\documentclass[sigconf]{acmart}
\AtBeginDocument{%
  \providecommand\BibTeX{{%
    \normalfont B\kern-0.5em{\scshape i\kern-0.25em b}\kern-0.8em\TeX}}}

\usepackage{enumitem}
\usepackage{ulem}
\usepackage{microtype}        
\usepackage{ccicons}          
\usepackage{multirow}
\usepackage{multicol}
\usepackage{color, colortbl}
\usepackage{makecell}
\usepackage{float}
\usepackage{balance}       
\usepackage{graphics}      
\usepackage[T1]{fontenc}   
\usepackage{txfonts}
\usepackage{mathptmx}
\usepackage{hyperref}
\usepackage{color}
\usepackage{booktabs}
\usepackage{xcolor}
\usepackage{textcomp}
\usepackage[T1]{fontenc}       
\usepackage[utf8]{inputenc}  
\usepackage[english]{babel}    
\usepackage{lmodern}
\setcopyright{acmcopyright}
\copyrightyear{2018}
\acmYear{2018}
\acmDOI{XXXXXXX.XXXXXXX}

\acmConference[SIGCSE TS '24]{Technical Symposium}{March 03--05,
  2024}{Woodstock, NY}
\acmPrice{15.00}
\acmISBN{978-1-4503-XXXX-X/18/06}

\begin{document}

\title[ChatGPT's strengths and weaknesses in solving UG CS questions]{ChatGPT in the Classroom: An Analysis of Its Strengths and Weaknesses for Solving Undergraduate Computer Science Questions}


\author{Ishika Joshi}
\email{ishika19310@iiitd.ac.in}
\affiliation{%
  \institution{IIIT Delhi}
  \city{New Delhi}
  \country{India}
}

\author{Ritvik Budhiraja}
\email{ritvik19322@iiitd.ac.in}
\affiliation{%
  \institution{IIIT Delhi}
  \city{New Delhi}
  \country{India}
}

\author{Harshal Dev}
\email{harshal19306@iiitd.ac.in}
\affiliation{%
  \institution{IIIT Delhi}
  \city{New Delhi}
  \country{India}
}

\author{Jahnvi Kadia}
\email{jahnvi21123@iiitd.ac.in}
\affiliation{%
  \institution{IIIT Delhi}
  \city{New Delhi}
  \country{India}
}

\author{M. Osama Ataullah}
\email{osama21127@iiitd.ac.in}
\affiliation{%
  \institution{IIIT Delhi}
  \city{New Delhi}
  \country{India}
}

\author{Sayan Mitra}
\email{sayan21142@iiitd.ac.in}
\affiliation{%
  \institution{IIIT Delhi}
  \city{New Delhi}
  \country{India}
}

\author{Harshal D. Akolekar}
\email{harshal.akolekar@iitj.ac.in}
\affiliation{%
  \institution{Dept of Mechanical Eng. \& School of AIDE}
  \city{IIT Jodhpur, Jodhpur}
  \country{India}
}

\author{Dhruv Kumar}
\email{dhruv.kumar@iiitd.ac.in}
\affiliation{%
  \institution{IIIT Delhi}
  \city{New Delhi}
  \country{India}
}

\renewcommand{\shortauthors}{Anonymous Authors}

\begin{abstract}

This research paper aims to analyze the strengths and weaknesses associated with the utilization of ChatGPT as an educational tool in the context of undergraduate computer science education. ChatGPT's usage in tasks such as solving assignments and exams has the potential to undermine students' learning outcomes and compromise academic integrity. This study adopts a quantitative approach to demonstrate the notable unreliability of ChatGPT in providing accurate answers to a wide range of questions within the field of undergraduate computer science. While the majority of existing research has concentrated on assessing the performance of Large Language Models in handling programming assignments, our study adopts a more comprehensive approach. Specifically, we evaluate various types of questions such as true/false, multi-choice, multi-select, short answer, long answer, design-based, and coding-related questions. Our evaluation highlights the potential consequences of students excessively relying on ChatGPT for the completion of assignments and exams, including self-sabotage. We conclude with a discussion on how can students and instructors constructively use ChatGPT and related tools to enhance the quality of instruction and the overall student experience.
\end{abstract}




\received{20 February 2007}
\received[revised]{12 March 2009}
\received[accepted]{5 June 2009}

\maketitle

\section{Introduction}\label{sec:intro}

One of the latest advancements in AI that has attracted a wide range of reactions is ChatGPT. 
ChatGPT is a language model trained by OpenAI that is based on the GPT-3.5\footnote{ChatGPT's free version uses GPT-3.5 while its paid version uses GPT-4.} architecture \cite{openai_introducing_2022}. It was made available to the public in November 2022 and since then it has attracted millions of users who are trying to use and test the AI tool \cite{shankland_why_2023}. Trained on a large dataset of internet-based text, ChatGPT is capable of producing text responses that resemble human-like language when provided with a prompt. Its capabilities extend to answering queries, engaging in diverse discussions, and creating original pieces of written work  \cite{taecharungroj_what_2023}. 

However, a sentiment of frenzy and fear has also been observed among professionals from various domains. ChatGPT has been feared to take away jobs of programmers, writers, specialists, educators, etc. \cite{noauthor_will_2023}. The influence ChatGPT can have on traditional learning and teaching academic practices is one such domain that has attracted a lot of debate and discussion \cite{becker2023ProsAndCons, Malinka2023Security, Daun2023Software, Denny2023CopilotCS1, Finnie-Ansley2022CS1, wermelinger2023Copilot, Savelka2023MCQAndCode, Reeves2023Parsons, finnie-ansley2023CodexCS2, Ouh2023Java, Cipriano2023GPT-3OOP, sarsa2022AutoGenerate, Leinonen2023CodeExplanation, Leinonen2023ExplainError, MacNeil2023CodeExplain, Balse2023Feedback}. Some people in academic circles have observed that students could use ChatGPT for cheating and plagiarism, but there are also others who argue that ChatGPT can be a beneficial tool for generating ideas and demonstrating responsible use of technology  \cite{noauthor_reactions_nodate}. 
Some students have expressed concern that such a tool could stifle their creativity and critical thinking skills \cite{the_learning_network_what_2023}. 

To cater to these rising concerns and the general uncertainty surrounding the implications and influence of ChatGPT in education, in this paper, we take a quantitative approach to analyze the perceived and debated threats of ChatGPT in academic contexts, particularly in the field of computer science. While the majority of existing research by the computing education community has concentrated on assessing the performance of Large Language Models (LLMs) in handling programming assignments, our study adopts a more comprehensive approach. Specifically, we evaluate various types of questions such as true/false, multi-choice, multi-select, short answer, long answer, design-based, and coding-related questions. 

More specifically, we aim to answer the following research questions in this paper:
\begin{itemize}[leftmargin=*]
    \item \textbf{Research Question 1:} What are the strengths and weaknesses of ChatGPT when answering various types of computer science questions? 
    \item \textbf{Research Question 2:} How can ChatGPT be constructively used by students and instructors to enhance their learning and teaching experience respectively?
\end{itemize}

To answer the above questions, we evaluate ChatGPT's capability in computer science across multiple topics, including core undergraduate courses, coding interview questions, and competitive examination questions. 
\section{Related Work}

ChatGPT has been widely praised for its ability to generate human-like responses, leading to its increased use in various industries, including academia. Several recent studies in the computing education community have examined ChatGPT's strengths and weaknesses from various viewpoints \cite{becker2023ProsAndCons, Malinka2023Security, Daun2023Software, Denny2023CopilotCS1, Finnie-Ansley2022CS1, wermelinger2023Copilot, Savelka2023MCQAndCode, Reeves2023Parsons, finnie-ansley2023CodexCS2, Ouh2023Java, Cipriano2023GPT-3OOP, sarsa2022AutoGenerate, Leinonen2023CodeExplanation, Leinonen2023ExplainError, MacNeil2023CodeExplain, Balse2023Feedback}.

Becker et al. \cite{becker2023ProsAndCons} discuss the various challenges and opportunities associated with computer science students and instructors using AI code generation tools such as OpenAI Codex, DeepMind AlphaCode, and Amazon CodeWhisperer. For instance, LLMs could be very helpful to instructors and students in generating high-quality learning material such as programming exercises, code explanations, and code solutions \cite{sarsa2022AutoGenerate}. At the same time, students may also indulge in unethical usage of LLMs for solving open-book assignments and exams. Similar challenges and opportunities have also been discussed in \cite{Denny2023CopilotCS1, Malinka2023Security, Daun2023Software}. A number of research studies have focused on evaluating how accurate are LLM models (such as OpenAI Codex, GPT-3, ChatGPT (GPT-3.5 and GPT-4)) in generating solutions for programming assignments in various computer science courses such as CS1 \cite{Denny2023CopilotCS1, Finnie-Ansley2022CS1, wermelinger2023Copilot, Savelka2023MCQAndCode, Reeves2023Parsons}, CS2 \cite{finnie-ansley2023CodexCS2, Savelka2023MCQAndCode}, object-oriented programming \cite{Ouh2023Java, Cipriano2023GPT-3OOP}, software engineering \cite{Daun2023Software}, and computer security \cite{Malinka2023Security}. These research studies showcase that LLMs are capable of generating reasonable solutions for a wide variety of questions albeit with varying accuracy. The accuracy depends on factors such as problem complexity and input prompt quality. 

Furthermore, multiple studies evaluate the ability of the LLMs to generate code explanations and compare the quality of these explanations with that of students \cite{Leinonen2023CodeExplanation, sarsa2022AutoGenerate, wermelinger2023Copilot, MacNeil2023CodeExplain}. Leinonen et al. \cite{Leinonen2023ExplainError} analyze how well can OpenAI Codex explain different error messages which a programmer may encounter while running a piece of code and how good are the corresponding code fixes suggested by Codex. This can be very helpful in debugging a program. Balse et al. \cite{Balse2023Feedback} investigate the potential of GPT-3 in providing detailed and personalized feedback for programming assessments which is otherwise not possible in a large class of students. This study finds that although the GPT-3 model is capable of correct feedback, it also generates incorrect and inconsistent feedback at times. Hence, its generated feedback must be verified by a human expert before it can be shared with students. 
\section{Research Contributions}
While prior research in the computing education community has concentrated on evaluating LLMs in the context of programming assignments in undergraduate computer science courses, our study focuses on  evaluating a wide variety of questions comprising true/false, multi-choice, multi-select, short answer, long answer, design-based, and coding-related questions. In our investigation, we focus on mid-term and end-term papers from four critical computer science subjects: data structures and algorithms, databases, operating systems, and machine learning. Additionally, we examine the Graduate Aptitude Test in Engineering (GATE), which comprises multiple-choice questions and assesses the knowledge of undergraduate/graduate students aspiring to pursue postgraduate programs in India. Finally, our assessment also includes full programming exercises. However, instead of evaluating the conventional CS1 or CS2 programming assignments, which have been subject to prior evaluations \cite{Denny2023CopilotCS1, Finnie-Ansley2022CS1, wermelinger2023Copilot, Savelka2023MCQAndCode, Reeves2023Parsons, finnie-ansley2023CodexCS2, Savelka2023MCQAndCode}, we concentrate on programming questions sourced from LeetCode. The LeetCode platform serves as a popular resource for practicing coding questions frequently encountered in interviews conducted by software companies. To summarize, our research endeavors to encompass a wide array of undergraduate computer science courses and diverse question types utilized to evaluate the proficiency of students in this field.

\section{Methodology}

\subsection{Research Design}
We utilize a quantitative research methodology to conduct a thorough analysis of ChatGPT's performance in response to questions posed from examinations undertaken by  undergraduate computer science students. Additionally, we also present a qualitative discussion to examine the types of questions accurately addressed by ChatGPT and the nature of errors it may encounter.
\subsection{Data Collection}
In order to comprehensively evaluate ChatGPT, we cover questions from three broad categories:
    \subsubsection{Core subjects in CS undergraduate curriculum:} We chose four subjects commonly found in a computer science undergraduate curriculum. The chosen subjects encompass three foundational courses in computer science: \textit{Data Structures and Algorithms (DSA)}, \textit{Operating Systems (OS)}, and \textit{Database Management Systems (DBMS)}. Additionally, we have included an important elective course on \textit{Machine Learning (ML)}, which currently stands as one of the most sought-after elective offerings. For each of these four subjects, we collected questions and solutions from well-established, renowned, and prestigious universities (MIT, Stanford, UC Berkeley, IITs), from different years to get a good collection of questions.
    \subsubsection{Graduate Aptitude Test in Engineering:} The Graduate Aptitude Test in Engineering (GATE) is a national-level entrance exam in India conducted jointly by the Indian Institute of Science and seven Indian Institutes of Technology (IITs). GATE scores are widely used for admission to postgraduate programs in engineering, as well as for direct recruitment to various public sector organizations and research institutions in India. Since tens of thousands of final year and graduated students give the GATE exam every year, we considered it appropriate to include this in our evaluation.
    \subsubsection{Programming Questions from LeetCode:} Leetcode is a popular platform for practicing coding interview questions commonly asked by companies during the software development and related hiring processes. For undergraduate CS students, solving Leetcode questions is a useful way to prepare for technical job interviews and develop problem-solving skills. Since there are already some studies on the evaluation of ChatGPT on programming exercises asked in courses such as CS1 and CS2, we focused on evaluating ChatGPT on programming questions from LeetCode as it covers a wide range of programming questions. A prompt stating \textit{"You are a computer science UG student preparing for technical interviews. Please answer the below questions"} was given for Category 2 questions, while no prompt was given for Category 1.

More specific details about the exact data sources as well as the number and type of questions for each of the above-mentioned categories are presented in Table \ref{tab:results}.


\definecolor{Gray}{gray}{0.9}
\renewcommand{\arraystretch}{1.1}
\begin{table*}[!ht]
    \centering
    \footnotesize
    \resizebox{0.9\textwidth}{!}{%
    \begin{tabular}{|c|c|c|l|}
        \hline
        \parbox{7em}{\centering\vspace{0.3\baselineskip}\textbf{Subject}\vspace{0.3\baselineskip}} & \textbf{Types of Questions} & \textbf{Number of Questions}& \textbf{Data Source} \\
        \hline
        \rowcolor{Gray}
        \multirow{1}{*}{\hfil \textbf{\parbox{7em}{\centering Data Structures and Algorithms}}} & \parbox{17em}{\vspace{.3\baselineskip} True/False, short answers, long answers, design-based and coding-based questions\vspace{.3\baselineskip}} & 107 & \parbox{20em}{\vspace{.3\baselineskip} \textbf{MIT:} Spring 2020, May 2012 final term papers \\ \textbf{UC Berkeley:} CS61BL Summer 2014, Spring 2018 mid-term papers\vspace{.3\baselineskip}} \\
        \multirow{1}{*}{\hfil \textbf{Operating Systems}} & \parbox{17em}{\vspace{.3\baselineskip} True/False questions with justification, short answers, long answers, design-based and coding-based questions\vspace{.3\baselineskip}} & 101 & \parbox{20em}{\vspace{.3\baselineskip} \textbf{Stanford University:} CS101 Spring 2018, CS140 Autumn 2007 \\\textbf{UC Berkeley:} CS162 Spring 2013 Mid Term, CS162 Spring 2013 End Term, CS162 Fall 2013 Mid Term, CS162 Spring 2017 3rd Mid Term \vspace{.3\baselineskip}} \\
        \rowcolor{Gray}
        \multirow{1}{*}{\hfil \textbf{\parbox{10em}{\centering Database Management Systems}}} & \parbox{17em}{\vspace{.3\baselineskip} MCQs, true/false, MSQs, and theory questions (Numericals, fill-in-the-blanks, reasoning-based questions, design-level questions) \vspace{.3\baselineskip}} & 108 & \parbox{20em}{\vspace{.3\baselineskip} \textbf{Stanford University:} 2020, 2021 term papers \vspace{.3\baselineskip}} \\
        \multirow{1}{*}{\hfil \textbf{Machine Learning}} & \parbox{17em}{\vspace{.3\baselineskip}  MCQs, MSQs, short and long-answer type theory questions and mathematical derivations\vspace{.3\baselineskip}} & 111 & \parbox{20em}{\vspace{.3\baselineskip} \textbf{UC Berkeley:} 2021, 2022 term papers \vspace{.3\baselineskip}} \\
        \rowcolor{Gray}
        \multirow{1}{*}{\hfil \textbf{GATE}} & \parbox{17em}{\vspace{.3\baselineskip} MCQs, fill-in-the-blanks for theoretical and numerical concepts\vspace{.3\baselineskip}} & 100 & \parbox{20em}{Random sampling of Archive Questions from 2001 to 2023}  \\
        \multirow{1}{*}{\hfil \textbf{LeetCode Coding}} & \parbox{17em}{\vspace{.3\baselineskip} Technical coding questions\vspace{.3\baselineskip}} & 118 & \parbox{20em}{\vspace{.3\baselineskip} \textbf{Category 1:} Subtopics of Data Structures and Algorithms \\ \textbf{Category 2:} Blind75 curated list of frequently asked Leetcode problems\vspace{.3\baselineskip}} \\
        \hline
    \end{tabular}}
    \caption{Dataset details used for ChatGPT's evaluation.}
    \label{tab:results}
    \vspace{-1.5\baselineskip}
\end{table*} 
\renewcommand{\arraystretch}{1}

\subsection{Evaluation Process}
We took one question at a time and provided it to ChatGPT as a prompt (along with any choices wherever applicable). We saved the response given by ChatGPT as its answer to this question. We measured the accuracy of ChatGPT by comparing each question's response with the correct solution (available online from the same source as the questions). Each response from ChatGPT was analyzed by authors using their domain expertise and categorized as correct, incorrect, or partially correct. ChatGPT's responses for GATE questions were categorized either correct or incorrect as all the questions were objective in nature. 
\section{Results}\label{sec:results}
\definecolor{Gray}{gray}{0.9}
\newcolumntype{C}{>{\centering\arraybackslash}p{1.2em}}
\renewcommand{\arraystretch}{1.4}
\begin{table*}[!ht]
    \centering
    \footnotesize
    \resizebox{1\textwidth}{!}{%
    \begin{tabular}{|Cr|CCC|CCC|CCC|CCC|CCC|CCC|c|}
        \hline
        \multicolumn{2}{|c|}{\multirow{2}{*}{\textbf{Subject}}} &
            \multicolumn{3}{c|}{\textbf{True/False}} &
            \multicolumn{3}{c|}{\textbf{Short/Long}} &
            \multicolumn{3}{c|}{\textbf{Coding}} & 
            \multicolumn{3}{c|}{\textbf{Design}} &
            \multicolumn{3}{c|}{\textbf{Numerical}} & 
            \multicolumn{3}{c|}{\textbf{MCQ/MSQ}} &
            \multirow{2}{*}{\textbf{Accuracy \%}}\\
            & & {\textbf{C}} & {\textbf{P}} & {\textbf{T}} & {\textbf{C}} & {\textbf{P}} & {\textbf{T}} & {\textbf{C}} & {\textbf{P}} & {\textbf{T}} & {\textbf{C}} & {\textbf{P}} & {\textbf{T}} & {\textbf{C}} & {\textbf{P}} & {\textbf{T}} & {\textbf{C}} & {\textbf{P}} & {\textbf{T}} & \\
            \hline
            \rowcolor{Gray}
            \multicolumn{2}{|c|}{\parbox{9em}{\centering\vspace{.3\baselineskip}\textbf{Data Structures and Algorithms}\vspace{.3\baselineskip}}} & 
                32 & 3 & 40 &
                15 & 6 & 28 & 
                10 & 0 & 11 & 
                14 & 2 & 20 & 
                4 & 3 & 8 &  
                & - &  & \textbf{70.1} \\
            \multicolumn{2}{|c|}{\parbox{9em}{\centering\vspace{.3\baselineskip}\textbf{Operating Systems}\vspace{.3\baselineskip}}} &
                15 & 1 & 20 & 
                27 & 1 & 32 & 
                4 & 6 & 15 &
                5 & 0 & 5 &
                8 & 9 & 29 &
                & - &  & \textbf{58.4} \\
            \rowcolor{Gray}
            \multicolumn{2}{|c|}{\parbox{10em}{\centering\vspace{.3\baselineskip}\textbf{Database Management Systems}\vspace{.3\baselineskip}}} &
                2&  1&  5&
                16&  4&  53&  
                & - &  &  
                & - &  &  
                11&  6&  28&  
                7&  3&  22& \textbf{33.4}\\
            \multicolumn{2}{|c|}{\parbox{10em}{\centering\vspace{.3\baselineskip}\textbf{Machine Learning}\vspace{.3\baselineskip}}} &
                & - &  & 
                30&  2&  38&
                & - &  &  
                & - &  &  
                11&  3&  21&  
                16&  0&  52& \textbf{51.4}\\
            \rowcolor{Gray}
            \multicolumn{2}{|c|}{\parbox{10em}{\centering\vspace{.3\baselineskip}\textbf{GATE}\vspace{.3\baselineskip}}} &
                &  -&  &  
                &  -&  &  
                &  -&  &
                &  -&  &  
                &  -&  & 
                49 & 0 & 100 & \textbf{49.0}\\
            \multirow{2}{*}{\parbox{7em}{\centering\vspace{.3\baselineskip}\textbf{LeetCode}\vspace{.3\baselineskip}}}
                & Cat. 1 &  & - &  &  & - &  & 26 & 22 & 48 &  & - &  &  & - &  &  & - &  & \textbf{54.2} \\
                & Cat. 2 &  & - &  &  & - &  & 65 & 0 & 70 &  & - &  &  & - &  &  & - &  & \textbf{92.8} \\
            \hline
            \multicolumn{2}{|c|}{\parbox{9em}{\centering\vspace{0.3\baselineskip}\textbf{Category-wise Accuracy \%}\vspace{0.3\baselineskip}}} &
            \multicolumn{3}{c|}{\textbf{75.4}} &
            \multicolumn{3}{c|}{\textbf{58.3}} &
            \multicolumn{3}{c|}{\textbf{53.8}} & 
            \multicolumn{3}{c|}{\textbf{76.0}} & 
            \multicolumn{3}{c|}{\textbf{39.5}} & 
            \multicolumn{3}{c|}{\textbf{41.3}} &
            \\
        \hline
    \end{tabular}
    }
    \caption{Subject and question-category breakdown and accuracy measure. (C: Correct, P: Partially correct, T: Total)}
    \label{tab:newEval}
    \vspace{-2\baselineskip}
\end{table*}
\renewcommand{\arraystretch}{1}


Table \ref{tab:newEval} provides us with a summary of the results we obtained after following the specified methodology. It presents both the subject-wise and category-wise results. All figures are in absolute numbers, unless specified. The total number of questions for each subject can be found in Table \ref{tab:results}.

\subsection{Accuracy Analysis}
ChatGPT has a mean accuracy of 56.9\% in terms of correctly answering questions across all subjects and all categories, implying that ChatGPT is indeed highly unreliable when it comes to answering computer science questions.

\noindent\textbf{Subject-wise Accuracy.}
ChatGPT's performance varies across different subjects in an undergraduate CS program. Our results show that ChatGPT is best suited to answer prompts that are coding based sourced from leetcode and have a context-setting prompt, as it achieved the highest accuracy of 92.8\%. ChatGPT was least accurate in answering questions from Database Management Systems, with an accuracy of 33.4\%. Further, it had an accuracy of 70.1\% for Data Structures and Algorithms, 58.4\% for Operating Systems, 51.4\% for Machine Learning, 49\% for GATE and 54.2\% for  LeetCode category 1. 

\noindent\textbf{Category-wise Accuracy.}
Our results highlight that ChatGPT is most accurate in answering questions that are Design based in nature, with an accuracy of 76\%. On the contrary, ChatGPT had a minimum accuracy of 39.5\% in answering numerical questions. Moreover, True/False questions had an accuracy of 75.4\%, 58.3\% for Short/Long, 53.8\% for coding-based and 41.3\% for MCQ/MSQ questions. 

\subsection{Insights}
Throughout the evaluation phase, multiple levels of observations were made, including prompt-based and subject-based observations. Upon completion, the major findings were combined and have been listed as follows. These observations were then further used to propose a set of recommendations for students and instructors when it comes to integrating ChatGPT into their academic workflows.

\begin{itemize}[leftmargin=*]

    \item ChatGPT has inconsistencies and tends to answer basic questions incorrectly, even if they can be solved by a direct formula. On the other hand, following an unpredictable behaviour, it has provided well framed answers for more difficult questions. An example of a basic fact-based question answered incorrectly is: \\
    \noindent \textbf{\small State True or False: Given a directed graph G = (V, E), run breadth-first search from a vertex $s \in V$ . While processing a vertex u, if some $v \in Adj+(u)$ has already been processed, then G contains a directed cycle.} \\
    This behaviour was repeatedly observed across different types of questions, and different subjects, especially, but not limited to, Data Structures and Algorithms, Operating Systems and Database Management Systems.
\\
    \item Prompting ChatGPT without laying down the context has a tendency to lower the response accuracy, and has often led ChatGPT to fixate on wrong parts of the question. Re-establishing the context causes ChatGPT to approach the question differently, and has lead to better results and even regenerated, correct answers \cite{Reeves2023Parsons}. An example of teh same is as follows: \\
    \noindent \textbf{\small What is the value of x at the end of this code? Assume that 3pixels.jpg has 3 pixels. Show your work for partial credit. x = 3; y = 7; img = new SimpleImage("3pixels.jpg"); for (pixel : img) \{ x = x + y; x = x + 1; y = 1; \} } \\
    This prompt was answered correctly once the context of the question, including the subject to which it belongs, was specified to ChatGPT.
\\    
    \item In majority of the incorrect answers for multiple-select questions, ChatGPT has provided the correct explanation and reached the correct answers but failed to select the option corresponding to the correct answer as the final output, resulting in it selecting the wrong answer. There were cases where it selected a certain option as the correct option but it misread the option content or reported completely made-up option content. For the following question:
    \\
        \noindent \textbf{\small "The preorder traversal sequence of a binary search tree is 30, 20, 10, 15, 25, 23, 39, 35, 42. Which one of the following is the postorder traversal sequence of the same tree?
(A) 10,20,15,23,25,35,42,39,30 (B) 15,10,25,23,20,42,35,39,30
(C) 15,20,10,23,25,42,35,39,30 (D) 15,10,23,25,20,35,42,39,30" } \\
It gave the order as \textbf{15, 23, 25, 10, 35, 42, 39, 30} and selected option (A) as the answer but this order does not match with option (A)'s content or any other option's content. 
\\
    \item When ChatGPT was provided with the prompt \textit{"You are computer science UG student preparing for technical interviews. Please answer the below questions"} for LeetCode Category 2 questions, we observed a drastic improvement in the response accuracy when compared to ChatGPT's performance for Category 1 questions. Further, it is observed that ChatGPT responses are easily susceptible to incorrect answers as well. For any given question, when further prompts are provided which contradict the previous response, ChatGPT immediately apologies, assumes our prompt to be correct and incorrectly modifies its answer.
\\

    \item When users give more details or context to their initial question, it has been observed that ChatGPT sometimes overlooks or doesn't take into account what was previously discussed. Instead, it focuses only on the most recent information provided by the user. This behavior can result in the model producing responses that are incorrect because it hasn't properly considered all the relevant information from the ongoing conversation. In essence, the model's inability to retain and integrate past context can lead to responses that don't align with the overall discussion, thus compromising the accuracy of its generated responses.
\\
    \item In various instances, when prompted to generate a revised response, ChatGPT consistently offered answers that were incorrect. What's more, these responses were accompanied by explanations that differed significantly from the original line of reasoning. This recurring pattern of behavior contributed to a growing sense of skepticism regarding the system's actual grasp of the logic being presented.
\\
    \item There is a higher accuracy reported for subjective and theoretical questions. ChatGPT seems to be performing worse on numerical-based questions where the answer needs to be precise with objective reasoning involved. For most correct theoretical answers, ChatGPT provided a very detailed explanation of the answer which made it easy to follow its logic. For example, take the question - 
\noindent \textbf{\small "There are 5 balls in a bag. Each ball is either red or blue. Let $(\theta)$ (an integer) be the number of blue balls. We want to estimate $(\theta)$,
so we draw 4 balls with replacement out of the bag, replacing each one before drawing the next. We get “blue,” “red,” “blue,”
and “blue” (in that order).
Assuming $(\theta)$ is fixed, what is the likelihood of getting exactly that sequence of colors (expressed as a function
of $(\theta)$)? " } \\
To this, it responded- 
\noindent \textbf{\small "Assuming that the draws are independent and the probability of drawing a blue ball is $(\theta)$/5 (since there are $(\theta)$ blue balls out of 5 total balls), the likelihood of getting the sequence of colors "blue, red, blue, blue" is given by the product of the probabilities of each individual draw:
P("blue, red, blue, blue" | $(\theta)$) = ($(\theta)$/5) * (1-$(\theta)$/5) * ($(\theta)$/5) * ($(\theta)$/5)
}\\
\\
    
    \item Solving basic mathematical problems was difficult for ChatGPT, rendering it to be unreliable in the cases of aptitude based problems. For example, while prompting ChatGPT questions from Machine Learning, it failed to implement standard calculations:  \\
    \noindent \textbf{\small Compute the determinant of the matrix A = [[22/5, -6/5], [-6/5, 13/5]]}\\
    ChatGPT gave the answer det(A) = 322/25 while the correct answer is 10.
\end{itemize}
\section{Discussion}
\subsection{Strengths \& Weaknesses}
\textbf{Variability in ChatGPT's accuracy and the need for prompt contextualization.} ChatGPT exhibits a notable disparity in accuracy across various subject domains when dealing with theoretical questions. In machine learning questions, ChatGPT achieved an accuracy of 74.5\% whereas its accuracy dipped to a mere 34.1\% for theoretical questions related to database management systems (DBMS). This overall subpar accuracy poses a significant challenge in positioning ChatGPT as a dependable assistant or guide for students and educators within academic settings. One key factor contributing to this variable accuracy is the imperative requirement for contextualization of queries within ChatGPT's framework \cite{Reeves2023Parsons}. Our study revealed a noteworthy improvement in accuracy when ChatGPT was presented with the prompt: 
\textit{"You are a computer science undergraduate (UG) student preparing for technical interviews. Please answer the questions provided."} Under such contextualized conditions, ChatGPT exhibited an exceptional accuracy of 92.8\%, with the majority of responses being entirely correct and the remainder partially correct. 
Hence, by offering additional prompts subsequent to the initial response, users can guide the model's output and refine the information provided based on their feedback, empowering users with greater control over ChatGPT's performance and refining the generated responses accordingly.

\noindent \textbf{Higher accuracy for subjective and theoretical questions.} ChatGPT's accuracy exhibits a notable decrease for single-choice-correct GATE questions compared to theoretical questions, such as in machine learning and other subjects (as previously discussed). This discrepancy highlights that ChatGPT exhibits a higher probability of delivering accurate responses to theoretical questions, particularly when the question involves a certain degree of subjectivity. 
In addition, LeetCode "easy" prompts had the highest acceptance rates, whereas "hard" prompts had the lowest. This implies that the nature of theoretical questions requiring less reliance on computational calculations allow ChatGPT to leverage its acquired knowledge rather than implementing logic and calculations. 
In situations where the user lacks academic knowledge of theoretical or subjective nature, ChatGPT has a higher probability of providing convincing and "believable" answer prompts \cite{kabir2023answers}.
This presents an opportunity of leveraging ChatGPT's capabilities for theoretical knowledge in scenarios where users already possess a certain degree of fluency in the subject matter. 

\noindent \textbf{Bias in ChatGPT's underlying language model}. The subpar performance of ChatGPT on GATE questions as compared to other set of questions can also be attributed to the inherent biases present in language models like ChatGPT, where certain groups or topics are enhanced due to the nature of the training datasets \cite{becker2023ProsAndCons}. As GATE is an examination specific to India, it is plausible that ChatGPT's training data inadequately represents GATE-specific content. Consequently, to enhance ChatGPT's utility across all regions, it becomes imperative to have a more expansive and inclusive corpus of training data.

\subsection{Recommendations for Students and Instructors}\label{sec:recommendations}

From our findings, we have come up with a set of recommendations that can help students and instructors to incorporate LLMs like ChatGPT in their workflow in order to support overall learning and performance of students. Similar recommendations have also been discussed by prior work which we cite as needed.

As shown in our results, ChatGPT does not consistently provide accurate explanations and answers to the questions, which might require prior expertise for verifying it's correctness. Therefore, for closed-book components such as in-class exams, quizzes, competitive examinations such as GATE etc., students must utilize the resources provided by the instructors as well as reliable online resources to grasp the subject matter. Once they have understood the subject matter, they can further use ChatGPT to generate practice questions for the exams \cite{becker2023ProsAndCons, sarsa2022AutoGenerate, Daun2023Software}. Students prioritize effective learning \cite{oh_goal_2019}, and immediate practice questions are essential to their learning \cite{shen_comparison_2019}. 
As shown by \cite{farrokhnia_swot_2023}, ChatGPT's strength lies in generating contextually sensitive responses. Using this, students can utilize contextualization to generate questions of personalized difficulty levels based on their understanding of the subject. Further, students could probe ChatGPT for a hint, if required \cite{becker2023ProsAndCons}.  Instructors can also use ChatGPT constructively by designing questions that require critical thinking and higher-order cognitive skills that cannot be easily solved by simply providing a prompt to ChatGPT \cite{Daun2023Software, Ouh2023Java, Cipriano2023GPT-3OOP}. Introducing open-ended assignments can allow students to show their creativity, and increase their understanding of the subject. As established by \cite{shen_comparison_2019}, students prefer to apply their learnings through practice questions. Instructors can use ChatGPT to generate quizzes and assessments in a gamified \cite{zhang_developing_2019}, non-monotonous manner in order to increase student engagement and learning.

In open-book components such as take-home assignments, homework, projects, etc., there lies a possibility for students to plagiarize responses from ChatGPT, which in turn lowers their academic integrity and hinders their learning \cite{becker2023ProsAndCons}. Our results show that ChatGPT cannot be relied upon due to the high variability in its accuracy. Hence, a recommended strategy for students is that students use ChatGPT as an assistant for initial ideation and write-ups and then build upon these through their own creativity and originality to fulfill the project requirements \cite{becker2023ProsAndCons}. This shall also positively influence the self-efficacy of the student, which is a major indicator of their level of understanding and engagement in the context of computer science courses \cite{sharmin_impact_2019}. Moreover, for take-home and open-book evaluation components, having questions that are precise and objective in nature might be a better way for instructors to test a student's knowledge compared to subjective questions. 
We saw in our results that ChatGPT provided detailed explanations but was often wrong in selecting specific answers.
This might result in a more balanced way of evaluation, where students are open to using tools like ChatGPT to assist their already existing fluency in the subject. Instructors can even try newer styles of evaluations where they give 
students a topic, and ask them to design a problem around that topic, with solutions and test cases (if applicable). A recent study used a similar methodology and observed that students showed an improvement in their performance \cite{kangas_does_2019}.  Students would be able to use ChatGPT as an assistant in problem generation, and verify their solutions through instructor-delivered in-class knowledge.

Studies have shown that students prefer to learn at their own pace \cite{shen_comparison_2019}. Some students find the class fast-paced, while the ones who are familiar with the coursework often tend to find classes slow-paced. Given the descriptive nature of explanations given by ChatGPT in our analysis, students can use ChatGPT to formulate their own outline and pace, rather than relying on the difficulty level of the class. This requires the student to have familiarity with the subject, as they can then verify the responses of ChatGPT and use it as a tool, rather than a replacement for the instructor.  

It is also important for students to be trained in asking the right kind of questions \cite{gupta_ChatGPT_2023} to ChatGPT as ChatGPT is sensitive to contextualisation, and the final answer of the prompt will depend entirely on the way it is asked by the user \cite{Reeves2023Parsons}.

\subsection{Limitations and Future Work}
\begin{itemize}[leftmargin=*]
    \item The present study evaluates ChatGPT's performance on a set of questions and uses this evaluation to understand the challenges and opportunities associated with its usage. The current analysis does not involve perspectives of students and teachers who have been using ChatGPT for their respective use cases. As part of our future work, we aim to collect and analyze the perspectives of students and teachers to gain deeper insights into the academic impact of ChatGPT.
    \item The primary finding of our current study unveils how ChatGPT demonstrates significant unreliability in generating accurate answers to provided questions, emphasizing the need for cautious utilization. We expect that ChatGPT's accuracy and performance will undergo enhancement over time, akin to the improvements observed in other AI models in the past. It will be intriguing to witness the point at which ChatGPT becomes a dependable tool in our repertoire.
    \item We acknowledge that some of the recommendations discussed in the paper are based on our interpretation of the analysis presented. As part of our future work, we plan to evaluate these recommendations using controlled experiments inside and outside the classroom to further establish their validity.
\end{itemize}
\section{Conclusion}
In this paper, we took a quantitative approach to demonstrate ChatGPT's high degree of unreliability in answering a diverse range of questions pertaining to topics in undergraduate computer science. Our analysis showed that students may risk self-sabotage by depending on ChatGPT to complete assignments and exams. Based on our analysis, we discussed the challenges, opportunities and recommendations for constructive use of ChatGPT by both students and instructors.
\bibliographystyle{acm}
\bibliography{chatgpt-1}

\appendix

\end{document}